\documentclass[applsci,article,accept,pdftex,moreauthors]{Definitions/mdpi} 

\firstpage{1} 
\pubvolume{1}
\issuenum{1}
\articlenumber{0}
\pubyear{2026}
\copyrightyear{2026}
\externaleditor{Reiner Creutzburg}
\datereceived{1 February 2026} 
\daterevised{30 April 2026}
\dateaccepted{3 May 2026} 
\datepublished{7 May 2026} 

\usepackage{pgfplots}
\usepackage{algorithm}
\usepackage{algpseudocode}

\Title{Few-Shot Network Intrusion Detection Using Online \linebreak  Triplet Mining}

\Author{Jack Wilkie 
 $^{1,}$*\orcidA{}, Hanan Hindy $^{2}$\orcidB{}, Christos Tachtatzis $^{1}$\orcidC{}, Miroslav Bures $^{3}$\orcidD{} and Robert Atkinson $^{3}$\orcidE{}}

\address{%
$^{1}$ \quad Department of Electronics and Electrical Engineering, University of Strathclyde, Glasgow G1 1XQ, UK
\\
$^{2}$ \quad Faculty of Computer and Information Sciences, Ain Shams University, Cairo 11566, Egypt
\\
$^{3}$ \quad Faculty of Electrical Engineering, Czech Technical University, 160 00 Prague, Czech Republic
}

\corres{Correspondence: jack.wilkie.2017@uni.strath.ac.uk}

\abstract{Network intrusion detection systems play a vital role in protecting networks by detecting malicious network traffic which can then be investigated by a cybersecurity operations centre. State-of-the-art approaches utilise supervised machine learning methods to train a classification model to recognise known cyberattacks; however, these models require a large labelled dataset to train and show poor performance when trained on smaller datasets. In an attempt to address this shortcoming, anomaly detection models learn the distribution of benign traffic and flag non-conforming traffic as malicious. While these methods do not require malicious examples to train, they suffer from high false-positive rates rendering them impractical. As a result, networks may be particularly vulnerable when there are insufficient labelled instances of a specific attack class to train an effective classifier. This often occurs in newly established networks or when previously unseen types of attacks emerge. To address this challenge, this work proposes the use of a triplet network, utilising online triplet mining and a KNN classifier, which is able to perform few-shot classification, enabling effective intrusion detection after being trained on a limited number of malicious examples. Various online triplet mining algorithms were explored and model design choices, such as the inference algorithm and optimised distance metrics, were compared and evaluated through a series of ablation studies. The final model was compared against other state-of-the-art approaches in few-shot binary and multiclass classification, where the proposed approach was found to be competitive with existing methods when trained on as little as 10 malicious samples of each class.}

\keyword{internet of things; network intrusion detection; machine learning; contrastive learning; few-shot; limited data; artificial intelligence; classification; CICDS2017; Lycos2017}

\begin{document}

\section{Introduction}
\label{sec:introduction}

Network intrusions continue to be a growing problem afflicting many sectors of the digital economy, including healthcare, finance, and manufacturing. The cyber landscape continues to grow and evolve with increasingly complex services and network technologies being used to support these various economic sectors; the deployment of Internet of Things (IoT) solutions is just one example. The use of an ever richer blend of services and technologies continues to grow the attack surface of Information Technology (IT) and Operational Technology (OT) networks alike. The predictable result is an escalating number of network intrusions. With the average data breach remaining undetected for 212 days and costing companies \$4.12 million~\cite{ibm_data_breach}, there is clearly an urgent need for effective network intrusion detection systems~(NIDSs).

NIDSs employ algorithms used to discriminate between benign network traffic, generated from everyday activities, and malicious traffic, generated by attackers~(using binary classification). Some NIDSs also identify the particular type of malicious action being performed~(using multiclass classification). When NIDSs identify malicious activity, the cybersecurity operations centre~(CSOC) is alerted to deal with the reported intrusion. Traditional NIDSs use a rules-based approach, which involves an expert hard-coding a series of rules for identifying malicious traffic. Unfortunately, this approach is unscalable given the increasing proliferation of cyber threats. A more scalable approach is to adopt supervised machine learning techniques whereby a classifier is trained on a range of benign and malicious traffic classes in order to subsequently differentiate new instances. To be effective, these methods require significant volumes of training data, which is expensive and time-consuming to obtain for new networks. Additionally, when a network is established, or a new~(zero-day) cyberattack emerges, supervised classifiers are ineffective until a large quantity of these attacks can be recorded, labelled and used to train the next generation of classifiers. During this period, networks remain vulnerable to this new class of attack. To combat this, researchers have proposed unsupervised classification approaches, which are able to detect outliers as novel traffic types, likely to be malicious. The disadvantage of this approach is the large number of false positives that can occur, often as a result of introducing a new service. This gives rise to the need for machine-learning-based approaches that are able to provide good performance when there are either no samples of the new attack class~(zero-shot learning), or limited numbers of examples~(few-shot learning). 

The primary objective of this work focuses on the latter, few-shot learning approach. By developing a few-shot intrusion detection method that maintains a low false-positive rate, this work aims to reduce the vulnerability window suffered by NIDSs after the emergence of a zero-day attack or establishment of a new network. To achieve this, the classifier must be able to learn from a limited number of malicious samples, despite the extreme class imbalance caused by the dominance of benign samples in the NIDS datasets.

Accordingly, a triplet network featuring online triplet mining is proposed as a viable solution to training machine-learning-based NIDSs with a low number of malicious training samples of each class. It can be trained on as few as 10 malicious samples per class, and is resistant to overfitting, whilst still achieving excellent classification performance when paired with a K-Nearest Neighbours~(KNN) classifier. The proposed model is then compared to leading supervised and anomaly-based models from the literature on NIDS datasets, achieving competitive performance in the lowest-data setting and outperforming all baseline models in most few-shot binary and multiclass classification settings while maintaining substantially lower false-positive rates. This marks an important step towards ML-based NIDSs being used in practical scenarios. The contributions of this work can be summarised as follows:

\begin{enumerate}
    \item A network intrusion detection system is proposed, which utilises the combination of online triplet mining and a KNN classifier and is able to outperform baseline models when trained in a few-shot classification regime.
   
    \item A novel evaluation procedure is designed, and the triplet network is compared to leading models from the literature in both few-shot binary and multiclass classification. The proposed model shows competitive performance with 10 malicious samples per class and improves across a variety of performance metrics over baseline models in most settings with 20 or more malicious samples per class.

    \item A series of ablation studies are performed to determine the optimal triplet mining and inference algorithms. Additionally, analysis is conducted to demonstrate how the proposed methodology is suited to classification in a limited data regime by intrinsically reducing the impact of overfitting.
   \end{enumerate}

The remainder of this paper is arranged as follows: Section~\ref{sec:related_works} provides an overview of the relevant literature and previous works in NIDSs given a limited data environment, including anomaly-based methods, few-shot learning methods and previously proposed contrastive models. Section~\ref{sec:triplet_network} introduces the proposed approach, consisting of a triplet network, trained using online triplet mining, and a KNN classifier, which is empirically evaluated using NIDS datasets in both few-shot binary and multiclass classification in Section~\ref{sec:experiments}. Section~\ref{sec:Ablations} performs a series of ablation studies, to evaluate various training procedures and inference algorithms. Finally, relevant discussions and conclusions are provided in Section~\ref{sec:discussion_and_conclusion}.

\section{Related Works}
\label{sec:related_works}

This section provides an overview of the existing literature which is pertinent to this work, beginning with a review of NIDSs in Section~\ref{subsec:related_nids}. Following this, Section~\ref{subsec:related_anomaly_detection} explores the application of anomaly detection algorithms to NIDSs and Section~\ref{subsec:related_contrastive_learning} details the use of contrastive learning for NIDSs, including existing few-shot classifiers.

\subsection{Network Intrusion Detection Systems}
\label{subsec:related_nids}

NIDSs aim to enhance a network’s cyber defence by monitoring traffic and classifying it as either authorised benign traffic or unauthorised malicious traffic, which can then be escalated to the CSOC. Traditionally, this has been achieved using signature-based detection systems, in which domain experts manually craft features that uniquely identify specific classes of malicious traffic. Potential attacks are detected by matching observed traffic against these predefined signatures. While this approach typically yields high-precision detection, the expert effort required to develop and maintain signatures renders such systems difficult to scale.

In response to the limitations of signature-based detection, supervised machine learning has emerged as a scalable solution for NIDSs. These methods leverage large labelled datasets of network traffic to infer decision boundaries between benign and malicious behaviour, enabling effective classification. Gradient-free models exemplify some of the simplest approaches to supervised machine learning in NIDSs. For example, decision trees (DTs) learn to classify traffic by recursively partitioning the feature space using interpretable, rule-based splits. Random Forests (RFs) extend this approach by aggregating multiple decision trees trained on randomised feature subsets, improving robustness and reducing overfitting. Support Vector Machines (SVMs) perform classification by learning a maximum-margin hyperplane that separates benign and malicious samples in a high-dimensional feature space. Collectively, these models have demonstrated effectiveness in intrusion detection tasks using tabular network traffic data.

Similarly, deep learning approaches have also found success in NIDSs. Initially, deep belief networks~\cite{Wei2019} produced competitive performance. However, these were later supplanted by Multi-Layer Perceptrons (MLPs), which are now more commonly used. While alternative network architectures have been proposed, they are not well suited to tabular data. For example, Convolutional Neural Networks~(CNNs)~\cite{Xiao2019} are sensitive to the order of tabular features in the input vector and do not model the relationship between all features in early layers. Furthermore, an equivalent network can be found using a regularised feedforward neural network. Similarly, Long Short-Term Memory~(LSTM) networks, initially developed for time series data, have been employed in NIDSs~\cite{CNN_LTSM}. However, their inherent design for time series analysis means that tabular features must be processed sequentially, which can diminish performance and lead to inefficient training for tabular tasks. 

While machine learning models have achieved state-of-the-art (SOTA) performance in network intrusion detection tasks and enabled more scalable NIDSs, they are not without limitations. In particular, both gradient-free and deep learning approaches typically rely on large volumes of labelled training data to achieve robust generalisation. This dependency introduces a critical vulnerability window in operational settings: newly deployed networks or emerging zero-day attacks lack sufficient historical data for model training, leaving systems exposed during the interval between network deployment or attack emergence and the accumulation of sufficient labelled data. 

\subsection{Anomaly Detection}
\label{subsec:related_anomaly_detection}

In an attempt to overcome the reliance of supervised machine learning models on large labelled datasets, anomaly detectors instead aim to learn the distribution (or a distributional likelihood proxy) of benign traffic, allowing malicious traffic to be identified as outliers at test time. As they are trained exclusively on benign traffic during training, anomaly detectors can, in principle, detect both known and zero-day attacks without requiring prior knowledge of specific malicious classes.

Statistical anomaly detection leverages distance metrics or probabilistic models to identify deviations from normal network behaviour, thereby flagging potentially malicious traffic. Distance-based methods construct a reference vector from a set of tabular features using summary statistics such as the mean or median, or more generally a learned centroid or distribution. Test samples are then assigned an anomaly score based on their Minkowski distance from this reference vector~\cite{doi:10.1137/1.9781611976236.18}. Extensions of this approach generalise to the Frobenius and Grassmannian distances~\cite{10.1007/978-3-319-70087-8_59}. Moving away from global statistic, neighbourhood distance metrics such as nearest neighbour distance and local outlier factor assign anomaly scores by estimating local densities~\cite{electronics9061022}. Discriminative statistical methods can also be adapted for anomaly detection; for example, one-class SVMs~\cite{Hanan_svm1} and isolation forests~\cite{10040395} represent the extensions of SVMs and tree-based models, respectively.

While classical statistical approaches are computationally efficient and interpretable, deep learning-based anomaly detectors are better suited to capturing complex, non-linear patterns in high-dimensional data. Reconstruction-based approaches train a neural network to reproduce its input from a compressed or corrupted representation, with the assumption that models trained on benign data will reconstruct normal traffic accurately, while  large reconstruction errors will be produced for anomalous samples. Autoencoders implement this paradigm by learning a low-dimensional latent representation through a bottleneck layer~\cite{doi:10.1126/science.1127647, 9659562}. Conversely, Sparse autoencoders favour the use of regularisation over a bottleneck layer~\cite{9239385}. Variations of the autoencoder, such as DAE-LR~\cite{nkashama2024deeplearningnetworkanomaly} and DUAD~\cite{li2021deep}, aim to improve anomaly detection performance through the use of various techniques such as additional regularisation techniques and iterative data filtering. However, DUAD is unlikely to be applicable to a few-shot learning task due to its reliance on iterative data filtering in a regime in which data is scarce.

\textls[-15]{Generative deep learning models such as Generative Adversarial Networks (GANs)~\cite{9219561}} and Variational Autoencoders (VAEs)~\cite{9017945} aim to directly model the distribution of the benign input traffic. These learned distributions can then be employed for anomaly detection either by generating synthetic samples to train discriminative classifiers, or by approximating likelihood estimates under the learned distribution.

Finally, hybrid approaches aim to achieve performance gains by parameterising statistical techniques. Deep Gaussian Mixture Models~(DAGMMs)~\cite{purohit2020deepautoencodinggmmbasedunsupervised} combine an autoencoder with a Gaussian mixture model~(GMM), where the autoencoder learns a low-dimensional representation of benign data and the GMM estimates the density of this representation to assign anomaly scores. Similarly, AutoSVMs~\cite{8463474} apply a one-class SVM in the autoencoder's representation space. The Deep Support Vector Data Descriptor~(Deep SVDD)~\cite{pmlr-v80-ruff18a} parameterises distance-based methods by learning a representation in which benign traffic is clustered around a centroid in representation space.

While anomaly detectors are able to detect malicious traffic without labelled training examples, they typically exhibit poor classification performance, resulting in false-positive rates which are far too high for use in practical systems~\cite{10.1007/978-3-319-13563-2_27}.

\subsection{Contrastive Learning}
\label{subsec:related_contrastive_learning}

Contrastive learning has recently emerged as a promising approach for mitigating the vulnerability window faced by network intrusion detection systems after the establishment of a new network or emergence of a zero-day attack, owing to its ability to train effective classifiers from limited labelled data. Unlike conventional supervised learning approaches that rely on absolute class labels, contrastive learning leverages relative similarity relationships between samples to learn discriminative feature representations that can generalise beyond the classes observed during training.

\begin{figure}[H]
     \isPreprints{\centering}{}
     \includegraphics[width= 5in]{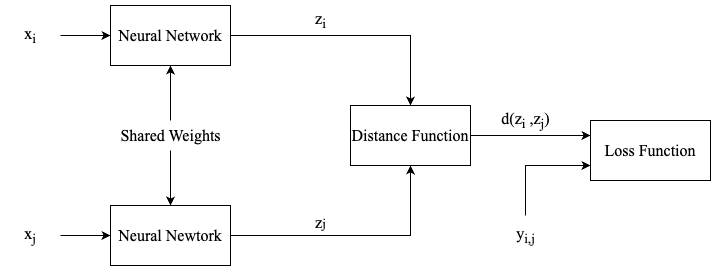} 
     \caption{Architecture of the Siamese network.\label{fig:siamese_diagram}}
\end{figure}

The simplest and most widely studied contrastive learning architecture is the Siamese network. As illustrated in Figure~\ref{fig:siamese_diagram}, the Siamese network consists of two identical neural networks with shared weights that process a pair of input samples \((x_i, x_j)\), where \(x_i,x_j \in \mathbb{R}^f\), in parallel. 

Training relies on a binary similarity label, $y_{i,j} \in \{0,1\}$, which indicates whether the pair of input samples are considered to be similar (\(y_{i,j} = 1\)) or dissimilar (\(y_{i,j} = 0\)). In supervised contrastive learning, samples of the same class distribution are regarded as similar, whilst sample pairs of differing class distributions are dissimilar. The contrastive loss function, defined in Equation~(\ref{contrastive-loss1}), minimises a distance metric, \(d: \mathbb{R}^{f_o} \times \mathbb{R}^{f_o}\!\to\mathbb{R}^+\) for output dimensionality \(f_o \in \mathbb{Z}^+\), typically the Euclidean distance, between embeddings of similar sample pairs while maximising the distance metrics up to a margin, \(m \in \mathbb{R}^+\), for dissimilar pairs. Here \(z := \phi(x) \in \mathbb{R}^{f_o}\) is used to represent the output vectors (known as embeddings) of the neural network \(\phi: \mathbb{R}^f \!\to \mathbb{R}^{f_o}\); the parameters \(\theta\) are omitted \linebreak  for simplicity.

\begin{linenomath}
\begin{equation}
L_{\mathrm{CL}}(x_i, x_j)
=
y_{i,j}\, d(z_i, z_j)^2
+
(1 - y_{i,j})\, \bigl(\max\{0,\, m - d(z_i, z_j)\}\bigr)^2
\label{contrastive-loss1}
\end{equation}
\end{linenomath}

As the contrastive loss directly optimises a geometric distance between sample embeddings, the output of the Siamese network can be interpreted as a learned geometric embedding space. In this space, the relative positions of samples encode semantic similarity, where embeddings of samples drawn from the same class distribution are encouraged to lie close together, while embeddings of samples drawn from different class distributions are separated by at least a margin. In an optimal embedded space, the smallest inter-class distance is larger than the greatest intra-class distance.

Once the Siamese network is trained, the resulting embedding space can be exploited for downstream tasks such as classification or similarity-based inference by comparing the distance of test samples to training embeddings of each class. This approach has been found to be robust when the Siamese network is trained on a limited training dataset. Previous works have applied Siamese networks to network intrusion detection, including few-shot learning, demonstrating their potential for detecting previously unseen attacks~\cite{hanan_developing_10.115/3437984.3458842,hanan_thesis}. However, these approaches relied on randomly sampled training and inference pairs, which can lead to slow convergence and sub-optimal classification performance. Subsequent work has explored alternative modalities, such as image-based representations of network traffic~\cite{Wang2021}, or other tasks such as zero-shot intrusion detection~\cite{hanan_leveraging_DBLP:journals/corr/abs-2006-15343}.

Triplet networks process an anchor sample $x_a \in \mathbb{R}^f$, a positive sample $x_p \in \mathbb{R}^f$ drawn from the same class distribution, and a negative sample $x_n \in \mathbb{R}^f$ drawn from a different class distribution using identical networks with shared weights. Training is guided by the triplet loss function, which enforces that the anchor is closer to the positive sample than to the negative sample by at least a margin $m \in \mathbb{R}^+$. By focusing on relative distance relationships rather than absolute similarity, triplet networks impose a less restrictive learning criterion that better preserves intra-class variability while maintaining inter-class separation. This formulation has been shown to produce more flexible and discriminative embeddings in other domains, such as face recognition~\cite{Schroff2015}.

A key limitation of Siamese networks in the context of network intrusion detection is their relatively strict learning constraint. By encouraging all samples from the same class to collapse toward a single point in the embedding space, Siamese networks may struggle to capture the significant intra-class variability present in network traffic, where benign behaviour and attack patterns often exhibit complex and partially overlapping distributions. In contrast, triplet networks optimise a relative distance constraint, requiring only that an anchor is closer to a positive sample than to a negative sample by a margin. This is a less restrictive objective, allowing the learned representation to preserve greater intra-class variability while still maintaining inter-class separation. This property is particularly beneficial in the few-shot intrusion detection setting considered in this work. Furthermore, the use of online triplet mining allows the model to exploit a larger number of informative training relationships than methods based on fixed pre-sampled pairs, which further contributes to improved performance.

In the context of intrusion detection, triplet-based contrastive learning has been explored for several adjacent purposes. RENOIR~\cite{Andresini2021} employed a triplet loss to accelerate convergence by constructing triplets using autoencoder reconstructions of benign and malicious traffic. Other work has used triplet losses for knowledge distillation, reducing model size while preserving classification performance for deployment in resource-constrained environments~\cite{Wang2022}. These studies demonstrate the potential of triplet networks to improve representation learning efficiency and robustness in network security tasks.

More recently, a number of studies have explored few-shot learning for network intrusion detection using alternative representation learning and meta-learning paradigms. A mutual centralised learning framework has been proposed to model bidirectional relationships between support and query samples, demonstrating strong performance in both binary and multiclass settings~\cite{xu2025fsmcl}. Prototypical capsule networks with attention mechanisms have been introduced to improve minority-class discrimination, while adaptive feature fusion strategies have been proposed to enhance representation quality in prototype-based few-shot learning~\cite{10.1371/journal.pone.0284632,bo2024featurefusion}. In addition, model-agnostic meta-learning approaches have been applied to intrusion detection to enable rapid adaptation to new attack classes, and class-incremental few-shot learning frameworks have been explored to allow intrusion detection systems to continuously incorporate newly observed attack types~\cite{10118898,cao2025fscil_ids}.

These studies highlight the growing interest in few-shot and limited-data intrusion detection; however, they primarily focus on prototypical, meta-learning, or incremental learning formulations. In contrast, this work builds on prior work that applied Siamese networks to few-shot intrusion detection~\cite{hanan_thesis}, by extending the contrastive learning framework to triplet networks. Additionally, several improvements are proposed, including the use of online triplet mining and a KNN classifier. By leveraging the increased flexibility of triplet-based learning, the proposed approach aims to better capture the complex structure of network traffic data and improve generalisation performance when only limited labelled samples are available.

\section{Proposed Approach}
\label{sec:triplet_network}
In this section, the triplet network is proposed for few-shot learning. An overview of the proposed system, which utilises a triplet network trained using online triplet mining, is given in Section~\ref{subsec:method_system_overview}. Section~\ref{subsec:method_balanced_sampling} begins to examine the proposed method by describing the balanced batch sampling approach used to train the model, ensuring that each batch contains valid triplets for the model to learn from. Next, the triplet network's architecture and training procedure are described in Section~\ref{subsec:method_triplet_network}. Finally, Section~\ref{subsec:method_inference} describes the KNN classification algorithm used to perform inference using the features learned by the \linebreak  triplet network.

\subsection{System Overview}
\label{subsec:method_system_overview}

\begin{figure}[H]
\isPreprints{}{
\begin{adjustwidth}{-\extralength}{0cm}
\centering
}
\includegraphics[width=\linewidth]{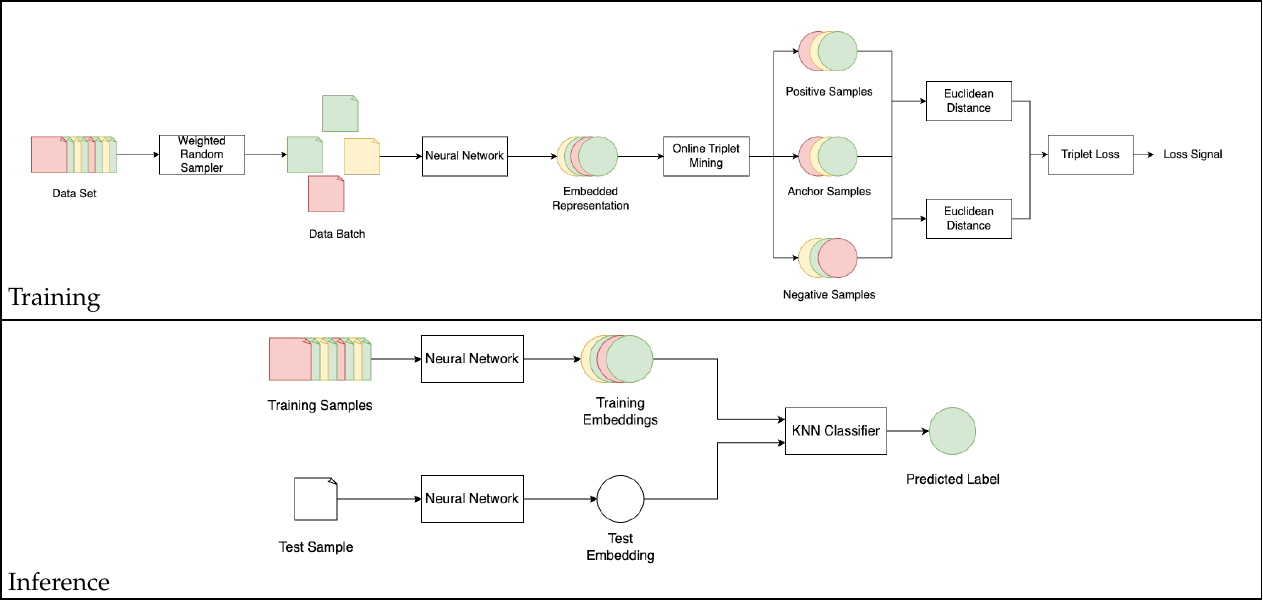}
\isPreprints{}{
\end{adjustwidth}
}
\caption{System overview of the proposed approach. (\textbf{Top}): Train-time configuration of the proposed approach. Training samples are converted to batches of embeddings using the neural network. Triplets are then dynamically assembled using online triplet mining and used to train the triplet network. (\textbf{Bottom}): Inference-time configuration of the proposed approach. The test sample is mapped to embedded space using the neural network, where it is classified using a KNN classifier by comparing its distance to train embeddings.}
\label{fig:system_overview}
\end{figure}

This work proposes the use of triplet networks and online triplet mining for network intrusion detection systems (NIDSs) in a few-shot learning scenario, extending prior research on the application of Siamese networks to similar tasks. The proposed approach, shown in Figure~\ref{fig:system_overview}, introduces several key modifications to improve performance and adaptability in the context of intrusion detection, such as the use of balanced batch sampling, online triplet mining, and a KNN classifier.

The train-time configuration of the proposed approach is illustrated in Figure~\ref{fig:system_overview}~(top). Initially, input batches of training data are sampled using class-balanced random sampling to ensure that triplets of each class type can be assembled. These batches are then mapped to an embedded representation using a neural network. Triplets can then be dynamically assembled from embedded batches using a process of online triplet mining and used to train the neural network using the triplet loss function. After training, the training set is passed through the neural network and the output embeddings are cached. Inference is then performed by calculating the test sample's embedding and applying a KNN classifier over the triplet network's embedded space, as shown in Figure~\ref{fig:system_overview}~(bottom).

\subsection{Balanced Batch Sampling}
\label{subsec:method_balanced_sampling}

As the majority of network flows originate from benign traffic, datasets exhibit extreme class imbalance; while benign traffic is abundant, the number of samples of malicious traffic is limited. This issue poses a significant challenge in this work, which aims to train models on as few malicious samples as possible. A further complication arises in training triplet networks using online triplet methods, as many batches of data contain only benign traffic and thus lack valid triplets for model training. Furthermore, previous works have reported slow convergence rates when training models using contrastive learning~\cite{Andresini2021}.

To address the aforementioned issues, a weighted random sampling strategy was used to assemble batches during model training. Instead of randomly shuffling the data and generating batches sequentially, or generating a fixed set of pairs prior to training as has been employed by prior work~\cite{hanan_thesis}, each sample was given a weight inversely proportional to the number of samples of its corresponding class. Batches were then randomly sampled from the training data according to the assigned weights. This ensured that the expected batch contained an even number of samples from each class, allowing for the 
formation of distinct triplets of each class combination. 

Concretely, given a training dataset
\(
\mathcal{D}_{\text{train}} := \{ (x_i, y_i) \mid x_i \in \mathbb{R}^f,\; y_i \in \mathcal{Y} \}_{i=0}^{N_{\text{train}} -1}
\),
containing \(N_{\text{train}} \in \mathbb{Z}^+\) training samples, where each sample \(x_i\) consists of \(f \in \mathbb{Z}^+\) tabular features representing a network flow, and
\(
\mathcal{Y} := \{0, \ldots, N_c\}
\)
denotes the set of absolute class labels, with \(y=0\) corresponding to benign traffic and \(y \neq 0\) corresponding to one of \(N_c \in \mathbb{Z}^+\) known malicious traffic classes, the sampling weights, \(w \in \mathbb{R}^+\), were calculated to be inversely proportional to the number of samples in the respective classes, as shown in Equation~(\ref{eq:weighted_sampling}). Here the set 
\(
\mathcal{D}_c := \{ (x_i,y_i) \in \mathcal{D}_{\text{train}} \mid y_i = c \},\, c \in \mathcal{Y}
\) 
is used to represent samples of the training data belonging to a given class distribution. Additionally, this work uses the notation \(y_i\)  to represent absolute class labels, while the notation \(y_{i,j}\) is used to notate relative similarity labels.
\begin{linenomath}
\begin{equation}
w_i := \frac{1}{|\mathcal{D}_{y_i}||\mathcal{Y}|},
\quad i = 0,\ldots,N_{\text{train}} -1,
\label{eq:weighted_sampling}
\end{equation}
\end{linenomath}

While weighted random sampling is a commonly used strategy to train MLPs on imbalanced data, it often results in sacrificed precision and increased false-positive rates to improve the recall of minority classes~\cite{Chawla_2002}; however, in this work the proposed triplet network, when combined with a KNN classifier, is found to be robust to this phenomenon, exhibiting both a high recall and  a low false-positive rate. 

\subsection{Triplet Network Model}
\label{subsec:method_triplet_network}

The proposed model is a triplet network trained to learn an embedding space in which samples from the same class are close and samples from different classes are separated by a margin. In contrast to conventional supervised classifiers that directly learn decision boundaries in the original feature space, the triplet network learns a representation that supports similarity-based inference in a limited data regime.

\begin{figure}[H]
\isPreprints{\centering}{}
\includegraphics[width= 5 in]{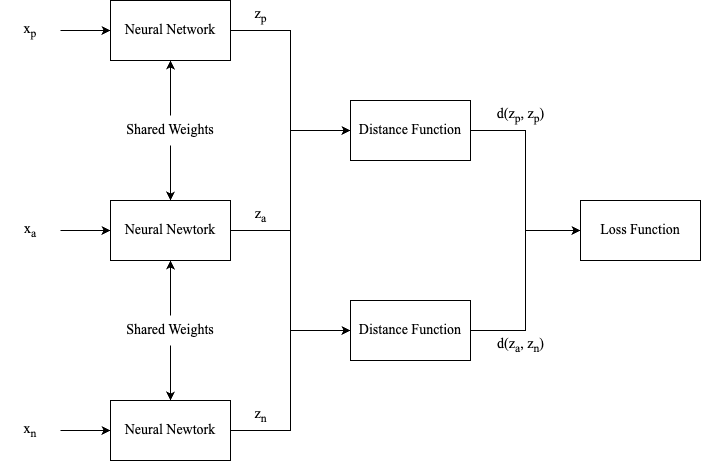} 
\caption{Architecture of the triplet network.\label{fig:triplet_diagram}}
\end{figure}

Triplet networks extend Siamese networks, which train on pairs of input samples, to instead train on triplets of input samples, allowing the network to learn the semantic intra-class structure. As shown in Figure~\ref{fig:triplet_diagram}, triplet networks can conceptually be viewed as three parallel neural networks, each with a shared set of weights.

The triplet loss function, shown in Equation~(\ref{eq:triplet_loss_1}), accepts three inputs: an anchor sample $x_a \in \mathbb{R}^f$, a positive sample $x_p \in \mathbb{R}^f$ and a negative sample \(x_n \in \mathbb{R}^f\) such that the anchor and positive samples are drawn from the same class distribution (\(y_a = y_p\)), while the negative sample is drawn from a different class distribution (\(y_a \neq y_n\)). Rather than enforcing absolute distance constraints in the embedding space, for the distance metric \(d: \mathbb{R}^{f_o} \times \mathbb{R}^{f_o}\!\to\mathbb{R}^+\), the triplet loss operates on relative distances, encouraging the distance between the anchor and positive samples to be smaller than the distance between the anchor and negative samples by a predefined margin, \(m \in (0, 1]\). This relative formulation promotes compact intra-class representations while simultaneously increasing inter-class separation, leading to more discriminative and well-structured embeddings.

\begin{linenomath}
\begin{equation}
L_{TL}(x_a,x_p,x_n) = \max\left\{0,\; d(z_a,z_p) - d(z_a,z_n) + m \right\}
\label{eq:triplet_loss_1}
\end{equation}
\end{linenomath}

A key limitation of prior few-shot NIDS contrastive approaches which utilise Siamese networks is the reliance on a fixed set of sampled pairs constructed before training~\cite{hanan_developing_10.115/3437984.3458842,hanan_leveraging_DBLP:journals/corr/abs-2006-15343}. This restricts training to a small subset of the available relationships and can lead to inefficient learning and poor classification performance if few informative pairs have been sampled. To address this, the proposed approach uses online triplet mining which, as shown in Figure~\ref{fig:system_overview}~(top), involves a reordering of operations: instead of explicitly selecting triplets in input space, a class-balanced batch is first passed through a single encoder network to produce embeddings. Informative triplets are then constructed within the batch based on distances in the embedding space.

Among online mining strategies, this work primarily adopts batch-all triplet mining, considering all valid triplets that can be formed within a batch. Let $x:=\{x_0,\ldots,x_{B-1}\}$ be a batch of size $B \in \mathbb{Z}^+$ with corresponding labels $y:=\{y_0,\ldots,y_{B-1}\}$. A triplet $(x_a,x_p,x_n)$ is valid if $y_p=y_a$ and $y_n\neq y_a$. The batch-all loss is computed as the mean triplet loss over all valid triplets in the batch where $N_t \in \mathbb{Z}^+$ is the number of valid triplets in the batch. Batch-all mining increases the amount of relational supervision extracted per batch, which is particularly beneficial in the few-shot setting where the total number of labelled malicious samples is small. Moreover, because only triplets that violate the margin contribute non-zero loss (Equation~(\ref{eq:triplet_loss_1})), the optimisation naturally focuses on informative triplets without requiring a fixed pre-sampled triplet set.

In each iteration, a batch is sampled using the balanced sampling strategy described in Section~\ref{subsec:method_balanced_sampling} to ensure that multiple classes are represented and that valid triplets exist. The batch is embedded via $\phi$, the batch-all mining objective is computed in embedding space, and gradients are backpropagated through the encoder using the batch-all triplet loss given in Equation~(\ref{eq:batch_all_triplet_loss}). After training, the encoder is used to embed training and test samples for inference, as described in Section~\ref{subsec:method_inference}.

\begin{linenomath}
\begin{equation}
L_{BA}(x, y) = \frac{1}{N_{t}} \sum_{a = 0}^{B-1} \sum_{p = 0 \atop y_p = y_a}^{B-1} \sum_{n = 0 \atop y_n \neq y_a}^{B-1} L_{TL}(x_a,x_p,x_n)\label{eq:batch_all_triplet_loss}
\end{equation}
\end{linenomath}

By training on all possible triplets within the dataset, the information available to the model during training is maximised, providing performance improvements over previously used contrastive approaches. Furthermore, the combination of online triplet mining and the hard margin in the triplet loss function allows for the model to learn from only the most difficult triplets (e.g., minority classes) from the set of all possible triplets. Thus, online triplet mining with the hinge-based triplet loss function allows for fast convergence during training and robustness to imbalanced training data.

\subsection{Inference}
\label{subsec:method_inference}

Unlike conventional classifiers trained with cross-entropy to directly model the conditional label distribution \(P(y|x)\), the proposed triplet network is trained to learn an embedded representation in which samples from the same class are closely clustered while samples from different classes are separated by a margin. Thus, inference can be performed by comparing the distance between the embedding of a test sample and embeddings of samples of each class distribution. While previous works applying Siamese networks to few-shot network intrusion compared test embeddings to randomly sampled training embeddings~\cite{hanan_thesis}, this work instead proposes the use of a KNN classifier to perform inference in the triplet network's embedded space.

Given a test sample $x_{\text{test}} \in \mathbb{R}^f$, its embedding $z_{\text{test}}:= \phi(x_{\text{test}})$ is computed and compared against embeddings of labelled training samples. The set of nearest neighbours to the test embedding, \(\mathcal{N}_k(z_{\text{test}})\), is then retrieved as the \(k \in \mathbb{Z}^+\) training embeddings with the least distance to the test embedding as shown in Equation~(\ref{eq:neighbours_calculation}).

\begin{linenomath}
\begin{equation}
\mathcal{N}_k(z_{\text{test}}) := \underset{\substack{\mathcal{S}\subset \mathcal{D}_{\text{train}} \\ |\mathcal{S}|=k}}{\arg\min} \sum_{(x_i,y_i)\in \mathcal{S}} d(z_{\text{test}},z_i)
\label{eq:neighbours_calculation}
\end{equation}
\end{linenomath}

The predicted class label, \(\hat{y} \in \mathcal{Y}\), is then obtained by majority vote as shown in Equation~(\ref{eq:knn_label_prediction}), where $\mathbf{1}\{\cdot\}$ is the indicator function which returns \(1\) when the condition is satisfied and \(0\) otherwise.

\begin{linenomath}
\begin{equation}
\hat{y} = \arg\max_{c \in \mathcal{Y}} \sum_{(x_i,y_i)\in \mathcal{N}_k(z_{\text{test}})} \mathbf{1}\{y_i = c\}
\label{eq:knn_label_prediction}
\end{equation}
\end{linenomath}

During training, weighted sampling is used to construct approximately class-balanced batches in order to ensure that informative triplets are present and to prevent the embeddings from being dominated by the majority (benign) class. This sampling strategy is beneficial for representation learning, ensuring that the triplet network learns class-balanced features; however, applying the same balancing at inference time would be undesirable in NIDSs because the true class distribution is highly skewed and benign traffic dominates in practice.

Using a KNN classifier for inference naturally reintroduces this skew: the neighbour search is performed over the (unbalanced) training set, so the expected composition of the retrieved neighbours reflects the prevalence of each class in the stored reference set. Consequently, the model can learn a discriminative embedding using balanced batches, while inference remains conditioned on the realistic class distribution, helping to maintain low false-positive rates.

More formally, if the embedding successfully clusters classes such that local neighbourhoods are class-consistent, then the expected number of neighbours from class $c$ among the $k$ retrieved neighbours satisfies and therefore depends on both class-conditional structure (captured by the embedding space) and the class prevalence in the reference set, as illustrated by Equation~(\ref{eq:expected_neighbours}).
\begin{equation}
\mathbb{E}[k_{\mathcal{C}_j}] \approx k \cdot P(y=\mathcal{C}_j \mid x_{\text{test}})
\label{eq:expected_neighbours}
\end{equation}

This provides an intuitive explanation for why the proposed architecture can achieve high recall for minority attack classes while preserving low false-positive rates in the presence of extreme class imbalance: unlike traditional supervised classifiers, which learn a class-balanced label distribution, the triplet network instead learns class-balanced features and then restores marginal label information at inference time through the KNN classifier, allowing for both high recall and low false-positive rates.

\section{Experimental Evaluation}
\label{sec:experiments}
In this section, the proposed triplet network is compared to models previously proposed in the literature. In Section~\ref{sec:dataset} the benchmark IDS datasets used for model evaluation are detailed and in Section~\ref{sec:evaluation_proceedure} the evaluation procedure is described. In Section~\ref{sec:binary_classification} the proposed model is compared to baseline models in a binary classification setting, including other contrastive models as well as anomaly-detection-based approaches. In Section~\ref{section:multiclass_classification} the proposed model is compared to baselines in a supervised classification scenario. Finally, the degree of overfitting seen in each of the evaluated models is given in Section~\ref{sec:overfitting}.

\subsection{Datasets}
\label{sec:dataset}

The proposed approach and baseline models were evaluated using two widely adopted benchmark datasets: CICIDS2017~\cite{cicids2017} and Lycos2017~\cite{lycos2017}. Although both datasets originate from the same underlying PCAP captures, CICIDS2017 was included to enable comparison with prior work~\cite{hanan_developing_10.115/3437984.3458842, hanan_leveraging_DBLP:journals/corr/abs-2006-15343}, which applies Siamese networks to a related few-shot intrusion detection task. In this work, an identical procedure was used to extract features from the PCAP files; however, the Siamese network was re-implemented and evaluated under the optimisation and evaluation procedure proposed in this work, allowing for a fair and controlled comparison across methods while isolating the impact of the proposed~approach.

Following this prior methodology, traffic in CICIDS2017 was represented using 28~bidirectional flow features across five representative classes: one benign class and four malicious classes (DoS Hulk, DoS Slowloris, FTP, and SSH). While these attack types are relatively well represented and can be distinguishable from benign traffic using flow-based features, they were selected primarily to ensure consistency with prior studies and to provide sufficient sample sizes for controlled experimentation. In contrast, attack types with very limited samples (e.g., SQL Injection) were excluded. Although such rare classes are important in real-world scenarios, their inclusion would introduce significant variance into the evaluation during both hyperparameter tuning and model evaluation on the test data. In particular, models that correctly classify only a small number of samples from these rare classes may achieve disproportionately high performance scores, despite not accurately capturing the overall class distribution. In this study, it was particularly important to minimise such sources of noise due to the low-data regime considered in this work, which can result in a significant amount of sampling noise in the training and validation data~splits.

The class-level composition of the CICIDS2017 dataset is shown in Table~\ref{cicds_data_composition}, highlighting the imbalance between benign and malicious traffic.

\begin{table}[H]
\caption{CICIDS2017 dataset composition. Class label 0 represents benign traffic, while labels 1--4 represent different malicious classes.}
\label{cicds_data_composition}
\isPreprints{\centering}{}
\centering
\begin{tabularx}{\textwidth}{CCC}
\toprule
\textbf{Class Label}    & \textbf{Class Name}     & \textbf{Num. Samples} \\
\midrule
0     & Benign & 248,607    \\
1     & DoS (Hulk)      & 14,427  \\
2     & DoS (Slowloris)       & 2840  \\
3     & FTP      & 5228 \\
4     & SSH       & 3627  \\
\bottomrule
\end{tabularx}
\end{table}

The Lycos2017 dataset~\cite{lycos2017} was additionally included to provide a more challenging and realistic evaluation setting. Lycos2017 addresses known issues in CICIDS2017, including incorrectly calculated features and labelling inconsistencies, resulting in a cleaner and more reliable dataset. Furthermore, Lycos2017 significantly increases the complexity of the task by providing 72 features and requiring discrimination across 12 classes. This introduces a more fine-grained classification problem, where distinctions between classes are less pronounced and decision boundaries are more difficult to learn. Additionally, the increased feature dimensionality exacerbates challenges associated with the curse of dimensionality, particularly in the low-data regime considered in this work. In few-shot settings, higher-dimensional feature spaces increase the risk of overfitting and reduce the reliability of distance-based and statistical estimates, making generalisation more difficult.

Following the same considerations as for CICIDS2017, minority classes with extremely few samples (Heartbleed and Web Attack—SQL Injection) were excluded from this study. Although these attack types are important in practice, their very small sample sizes do not allow for sufficient representation across training and evaluation splits. Including such classes would introduce additional sources of evaluation noise due to the introduction of additional sampling noise in the test data splits.

As a result, Lycos2017 serves as a complementary benchmark that better reflects real-world deployment scenarios, where attacks may be more diverse and less easily separable from benign traffic and other malicious traffic distributions. The class-level composition of the Lycos2017 dataset is shown in Table~\ref{lycos_data_composition}.

\begin{table}[H]
\caption{Lycos2017 dataset composition. Class label 0 represents benign traffic, while labels 1--11 represent different malicious classes.}
\label{lycos_data_composition}
\isPreprints{\centering}{}
\centering
\begin{tabularx}{\textwidth}{CCC}
\toprule
\textbf{Class Label}    &  \textbf{Class Name}     & \textbf{Num. Samples} \\
\midrule
0   & Benign    & 1,341,169 \\
1   & Port Scan    & 160,106 \\
2   & DoS (Hulk)    & 158,988 \\
3   & DDoS   & 956,83 \\
4   & FTP    & 8006 \\
5   & DoS (Golden Eye)    & 6765 \\ 
6   & SSH    & 5918 \\
7   & DoS (Slowloris)    & 5674 \\
8   & DoS (Slow HTTP Test)    & 4866 \\
9   & Web Attack (Brute Force)    & 1360 \\
10   & Botnet    & 735 \\
11   & Web Attack (XSS)    & 661 \\
\bottomrule
\end{tabularx}
\end{table}

\subsection{Evaluation Procedure}
\label{sec:evaluation_proceedure}

To evaluate the performance of the models across a range of limited dataset sizes, a bespoke model selection and evaluation procedure was employed. As described in \linebreak  Algorithm~\ref{alg:randomSamplingCV}, the dataset was initially partitioned into 50\%/50\% training and test splits, ensuring an equal number of samples from each class in both sets. A 50\%/50\% split was adopted to maximise the number of test samples in order to minimise evaluation noise, as a larger test set ensures that performance is evaluated over a more representative sample of each class, reducing sensitivity to specific minority class samples. The use of a relatively small training set is appropriate in this setting, as the few-shot regime restricts the number of training samples used per class and does not utilise the full training dataset. Feature engineering or selection was not performed in this work; instead, the PCAP parser used in prior studies~\cite{hanan_developing_10.115/3437984.3458842, hanan_leveraging_DBLP:journals/corr/abs-2006-15343} was used to extract 28 bidirectional flow features from the CICIDS2017 dataset, and the features contained in the provided CSV file were used for the Lycos2017~dataset.

\begin{algorithm}[H]
\caption{Model Selection and Evaluation with Random Sampling and Cross-Validation}
\label{alg:randomSamplingCV}
\begin{algorithmic}[1]

\Require Dataset $D$ with classes $C$, where each class is split 50/50 into training and testing sets.

\State Parameters:
\State \quad Number of subsets $S$
\State \quad Number of benign samples $N_B$
\State \quad Number of samples per malicious class $N_M$
\State \quad Number of folds $K$ in cross-validation

\For{each subset $s$ in $1$ to $S$}

\State Randomly sample $N_B$ benign instances
\State \quad and $N_M$ instances of each malicious class from the training set
\State \quad to form subset $D_s$

\State Initialize random search over the search space
\State \quad using $K$-fold cross-validation on $D_s$

\State Select the model $M_s$ with the highest mean F1 score across the $K$ folds.

\State Train $M_s$ on the entire subset $D_s$

\State Evaluate the performance of $M_s$ on the test set

\EndFor

\Ensure Performance metrics for each model $M_s$
\State \quad evaluated on the test set

\end{algorithmic}
\end{algorithm}

In order to control for sampling noise in the training data, which can cause significant variability in model performance when training on limited samples, the mean results across 10 independent hyperparameter optimisation and training runs are reported in this study. To accommodate this, the training data split was subdivided into \(S := 10\) subsets, each containing \(N_B := 10{,}000\) benign samples and \(N_M \in \{10,20,40,80,160\}\) samples uniformly sampled from the training data. Here, the low number of malicious samples was used to simulate the few-shot learning scenario investigated in this work. Conversely, a large number of benign samples was used to replicate a practical scenario where benign traffic is typically abundant, while malicious samples are limited. Additionally, the use of many benign samples allows for a more realistic marginal label distribution as datasets are often dominated by benign traffic.

Hyperparameter selection was performed separately for each of the \(S\) training subsets using five-fold cross-validation (\(K := 5\)), with folds stratified by class. For gradient-based models, hyperparameters were selected using 200 iterations of random search over predefined ranges, while for gradient-free models it was computationally feasible to perform an exhaustive grid search instead. The same tuning and evaluation procedure was applied to the proposed triplet network and all baseline models, with the exception that anomaly detection models were tuned using only benign training data, with malicious training data being instead added to the validation data split, allowing for better-tuned hyperparameters when compared to supervised approaches. To ensure fair comparison between models, the same held-out test set was used throughout. The full hyperparameter search space is reported in Appendix~\ref{app:hyperparameters}.

After the hyperparameter search, the best performing configuration was trained on the entirety of the training data subset and then evaluated on the unified test set. This was repeated for each training subset, with the mean results across all \( S \)  hyperparameter optimisation and training runs being reported. By conducting multiple hyperparameter searches and evaluations across different training splits, the results better capture variability due to sampling a limited training dataset. 
 
In this study, gradient-based models were trained for 200 epochs using the AdamW optimiser~\cite{loshchilov2019decoupledweightdecayregularization} with cosine annealing being applied to the learning rate. Weight decay was used as regularisation, with the strength being treated as a hyperparameter. Weighted random sampling was used to generate class-balanced training batches, while the full unbalanced training set was used for KNN evaluation. Input features were normalised using z-score normalisation, with the normalisation statistics being calculated independently for each training subset and cross-validation fold.

All experiments in this study using gradient-based models were run on an NVIDIA GeForce RTX 3090 GPU using 5 concurrent processes. An Intel Xeon W-2255 CPU was used for gradient-free models. The random seeds and software versions used for data partitioning and hyperparameter searching are provided in Appendix~\ref{app:hyperparameters}.

\subsection{Binary Classification}
\label{sec:binary_classification}

In binary classification, where the model must predict whether unknown traffic is benign or malicious, the proposed triplet network was compared to autoencoders~\cite{hanan_svm2}, DAE-LR~\cite{nkashama2024deeplearningnetworkanomaly} and one-class SVMs~\cite{Hanan_svm1}, representing anomaly-detection-based approaches, and RENOIR~\cite{Andresini2021}, which was used a state-of-the-art contrastive learning baseline. The macro F1 score, balanced recall, balanced precision, and false-positive rate for the proposed model and baselines when trained on varying amounts of malicious traffic are given in Table~\ref{tab:binary_results}. It should be noted that while the one-class SVM and autoencoder are trained only on benign samples, the malicious samples were used exclusively for validation and so they should have better-tuned hyperparameters when more malicious samples are available.

\begin{table}[H]
   \centering
   \caption{Comparison of the proposed triplet network with several baseline models across several datasets in binary classification. The models are compared across a variety of training set sizes. Bold values indicate the best performance for each metric, with lower values preferred for false-positive rate.} 
   \label{tab:binary_results}
\small
\begin{adjustwidth}{-\extralength}{0cm}
\begin{tabularx}{\fulllength}{LLCCCCCCCC}
\toprule
\multirow{2}{*}{\textbf{Samples}} & \multirow{2}{*}{\textbf{Model}} & \multicolumn{4}{c}{\textbf{CICIDS2017}} & \multicolumn{4}{c}{\textbf{Lycos2017}} \\
 & & \textbf{F1-Score} & \textbf{Recall} & \textbf{Precision} & \textbf{FP Rate} & \textbf{F1-Score} & \textbf{Recall} & \textbf{Precision} & \textbf{FP Rate} \\
\midrule

\multirow{5}{*}{10} & Triplet & \textbf{0.9508} & \textbf{0.9180} & \textbf{0.9910} & \textbf{0.0001} & 0.9870 & 0.9822 & \textbf{0.9921} & \textbf{0.0015} \\
& RENOIR & 0.8620 & 0.8782 & 0.8547 & 0.031708 & \textbf{0.9876} & \textbf{0.9879} & 0.9872 & 0.006618 \\
& AE & 0.8351 & 0.7886 & 0.9326 & 0.007858 & 0.7513 & 0.7217 & 0.8690 & 0.019926 \\
& DAE-LR & 0.8280 & 0.7662 & 0.9540 & 0.002579 & 0.7427 & 0.7073 & 0.8787 & 0.011905 \\
& SVM & 0.3179 & 0.2698 & 0.4203 & 0.504890 & 0.2734 & 0.2510 & 0.3003 & 0.499617 \\
\midrule

\multirow{5}{*}{20} & Triplet & \textbf{0.9562} & \textbf{0.9276} & \textbf{0.9902} & \textbf{0.0004} & \textbf{0.9899} & 0.9875 & \textbf{0.9924} & \textbf{0.0026} \\
& RENOIR & 0.8512 & 0.8643 & 0.8615 & 0.026064 & 0.9879 & \textbf{0.9898} & 0.9861 & 0.008017 \\
& AE & 0.7926 & 0.7457 & 0.9201 & 0.006597 & 0.7636 & 0.7316 & 0.8632 & 0.024317 \\
& DAE-LR & 0.8504 & 0.7891 & 0.9600 & 0.002455 & 0.7792 & 0.7430 & 0.8895 & 0.013361 \\
& SVM & 0.3177 & 0.2695 & 0.4202 & 0.505430 & 0.2733 & 0.2509 & 0.3003 & 0.499817 \\
\midrule

\multirow{5}{*}{40} & Triplet & \textbf{0.8613} & 0.8433 & \textbf{0.8841} & \textbf{0.0001} & \textbf{0.9916} & 0.9913 & \textbf{0.9920} & \textbf{0.0039} \\
& RENOIR & 0.7950 & \textbf{0.9141} & 0.7647 & 0.119111 & 0.9903 & \textbf{0.9931} & 0.9875 & 0.007845 \\
& AE & 0.8460 & 0.7993 & 0.9246 & 0.008601 & 0.7940 & 0.7750 & 0.8542 & 0.043376 \\
& DAE-LR & 0.8696 & 0.8178 & 0.9542 & 0.004178 & 0.8194 & 0.7833 & 0.8951 & 0.018561 \\
& SVM & 0.3179 & 0.2698 & 0.4203 & 0.504874 & 0.2733 & 0.2508 & 0.3002 & 0.499884 \\
\midrule

\multirow{5}{*}{80} & Triplet & \textbf{0.9609} & 0.9343 & \textbf{0.9920} & \textbf{0.0002} & \textbf{0.9910} & 0.9927 & \textbf{0.9895} & \textbf{0.0062} \\
& RENOIR & 0.8632 & \textbf{0.9381} & 0.8322 & 0.064756 & 0.9905 & \textbf{0.9940} & 0.9872 & 0.008326 \\
& AE & 0.8180 & 0.7817 & 0.8797 & 0.015057 & 0.8843 & 0.9043 & 0.8733 & 0.084521 \\
& DAE-LR & 0.8785 & 0.8460 & 0.9243 & 0.010217 & 0.8441 & 0.8237 & 0.8918 & 0.035378 \\
& SVM & 0.3185 & 0.2704 & 0.4205 & 0.503456 & 0.2734 & 0.2510 & 0.3003 & 0.499488 \\
\midrule

\multirow{5}{*}{160} & Triplet & \textbf{0.9605} & 0.9398 & \textbf{0.9841} & \textbf{0.0019} & \textbf{0.9926} & \textbf{0.9947} & \textbf{0.9906} & \textbf{0.0058} \\
& RENOIR & 0.8275 & \textbf{0.9416} & 0.7789 & 0.089986 & 0.9904 & 0.9942 & 0.9867 & 0.008761 \\
& AE & 0.8489 & 0.8700 & 0.8316 & 0.037868 & 0.8554 & 0.8899 & 0.8392 & 0.124640 \\
& DAE-LR & 0.8959 & 0.8806 & 0.9131 & 0.014502 & 0.8411 & 0.8526 & 0.8366 & 0.092028 \\
& SVM & 0.3184 & 0.2700 & 0.4203 & 0.503287 & 0.2737 & 0.2514 & 0.3005 & 0.498752 \\
\bottomrule
\end{tabularx}
\end{adjustwidth}
\end{table}

On the CICIDS2017 dataset, the triplet network outperformed all models on F1 score and precision. It was outperformed by RENOIR on the recall metric for larger numbers of training examples; however, this comes at the cost of prohibitively high false-positive rates: RENOIR produced false-positive rates two orders of magnitude above the triplet network, rendering it impractical to deploy. On the Lycos2017 dataset, the triplet network had the highest F1 score in all cases except at very low numbers of samples (10 malicious samples per class); in this case, RENOIR demonstrated marginally superior performance, but once again this came at the cost of significantly higher false-positive rates. Additionally, the triplet network maintained the best precision and false-positive rate across all training dataset sizes.

\subsection{Multiclass Classification}
\label{section:multiclass_classification}

In multiclass classification, where the model must distinguish between benign traffic and different classes of intrusion, the proposed triplet network was compared to the Siamese network~\cite{hanan_thesis}, which represents the current SOTA in few-shot classification without retraining or performing few-shot classification on a subset of malicious classes whilst training on large amounts of data from others. Additionally, an MLP~\cite{lycos2017} trained using the cross-entropy loss function was used as a baseline to compare the proposed triplet network to traditional non-contrastive approaches. Finally, a KNN classifier~\cite{8525522} was selected to compare the discriminative power of the embedded representations learned by the triplet network with that of the original feature space. As the proposed approach uses the KNN for inference, it is expected that performance improvements shown over the KNN would extend to an arbitrary classifier trained on the triplet network's embeddings instead of the original features. While RENOIR shows competitive performance in binary classification, it is not readily adapted to a multiclass classification problem. This is because inference is performed by comparing two distances (the distance between the sample and the benign reconstruction, $d(x, \bar{x_b})$, and that between the sample and the attack reconstruction, $d(x, \bar{x_a})$), producing a binary label corresponding to the smaller of the two distances. Due to this limitation, RENOIR is not used as a baseline model in this section. 

The performance results for each model on the CICIDS2017 and Lycos2017 datasets are given in Table~\ref{tab:multiclass_results}. On the CICIDS2017 dataset the triplet network needed at least 40~samples of each class to learn an improved representation of the data; this can be seen by its improved F1 score over the KNN above this threshold. On the Lycos2017 dataset the triplet network achieved the highest F1 score across all training dataset sizes. The triplet network's improved performance on the Lycos2017 dataset is likely due to the increased complexity of the task: on the CICIDS2017 dataset the model must discriminate between five classes using 28 features, whereas on the Lycos2017 dataset there are 12 classes and 72 features. In this more difficult setting, the KNN is less able to provide accurate predictions without the training step provided by the triplet network. This suggests that the triplet network is more suited to more complex tasks and realistic scenarios than gradient-free approaches. While the MLP and Siamese network performed well in recall across both datasets, they had low precision and false-positive rates, which were orders of magnitude higher than those of the triplet network, making them less viable solutions.

\begin{table}[H]
\caption{Comparison of the proposed triplet network with several baseline models across several datasets in multiclass classification. Bold values indicate the best performance for each metric, with lower values preferred for false-positive rate}.
\label{tab:multiclass_results} 
\small
\begin{adjustwidth}{-\extralength}{0cm}
\isPreprints{\centering}{}
\centering
\begin{tabularx}{\fulllength}{CCCCCCCCCC}
\toprule
\multirow{2}{*}{\textbf{Samples}} & \multirow{2}{*}{\textbf{Model}} & \multicolumn{4}{c}{\textbf{CICIDS2017}} & \multicolumn{4}{c}{\textbf{Lycos2017}} \\
 & & \textbf{F1-Score} & \textbf{Recall} & \textbf{Precision} & \textbf{FP Rate} & \textbf{F1-Score} & \textbf{Recall} & \textbf{Precision} & \textbf{FP Rate} \\
\midrule

\multirow{4}{*}{10} & 
Triplet & 0.8757 & 0.7967 & 0.9916 & 0.000272 & \textbf{0.7802} & 0.7576 & \textbf{0.8240} & \textbf{0.001476}
\\

& Siamese & 0.6312 & 0.8060 & 0.6283 & 0.208661 &  0.6416 & 0.7707 & 0.6155 & 0.155338 
\\

&  MLP & 0.7892 & \textbf{0.8211} & 0.7870 & 0.023389 &  0.7615 & \textbf{0.8516} & 0.7535 & 0.012975 
\\ 

& KNN & \textbf{0.8825} & 0.8070 & \textbf{0.9928} & \textbf{0.000264} & 0.7582 & 0.7440 & 0.7876 & 0.001542
\\
\midrule

\multirow{4}{*}{20} & 
Triplet & 0.8514 & 0.7817 & 0.9511 & \textbf{0.000204} & \textbf{0.8119} & 0.8178 & \textbf{0.8127} & 0.002353 
\\

& Siamese & 0.6735 & 0.8127 & 0.6512 & 0.154800 & 0.6631 & 0.8201 & 0.6289 & 0.153519
\\

&  MLP & 0.7892 & \textbf{0.8211} & 0.7870 & 0.023389 &  0.7776 & \textbf{0.8869} & 0.7587 & 0.013883 
\\

& KNN & \textbf{0.8873} &0.8148 & \textbf{0.9905} & 0.000246 &  0.8024 & 0.8133 & 0.8012 & \textbf{0.002075}
\\

\midrule
\multirow{4}{*}{40} & 
Triplet & \textbf{0.8957} & 0.8260 & \textbf{0.9917} & \textbf{0.000151} & \textbf{0.8346} & 0.8755 & \textbf{0.8116} & 0.003305 
\\

& Siamese & 0.6431 & 0.8011 & 0.6597 & 0.254220 & 0.6628 & 0.8584 & 0.6227 & 0.086839 
\\

&  MLP & 0.7892 & 0.8211 & 0.7870 & 0.023389 &  0.7887 & \textbf{0.9054} & 0.7657 & 0.014829 
\\

& KNN & 0.8947 & \textbf{0.8268} & 0.9875 & 0.000375 & 0.8218 & 0.8581 & 0.8007 & \textbf{0.002714} 
\\

\midrule\multirow{4}{*}{80} & 
Triplet & \textbf{0.9008} & \textbf{0.8365} & \textbf{0.9875} & \textbf{0.000312} & \textbf{0.8343} & 0.9156 & \textbf{0.8029} & 0.005954 
\\

& Siamese &0.7023 & 0.8323 & 0.6695 & 0.118468 & 0.6775 & 0.8677 & 0.6307 & 0.074649
\\

&  MLP & 0.7892 & 0.8211 & 0.7870 & 0.023389 &  0.7870 & \textbf{0.9212} & 0.7448 & 0.014582
\\ 

& KNN & 0.8983 & 0.8334 & 0.9857 & 0.000412 &  0.8268 & 0.8777 & 0.7980 & \textbf{0.003405 }
\\

\midrule

\multirow{4}{*}{160} & 
Triplet & \textbf{0.9035} & 0.8469 & 0.9776 & 0.000615 &  \textbf{0.8446} & 0.9236 & \textbf{0.8109} & 0.005641
\\

& Siamese & 0.7519 & \textbf{0.8495} & 0.7246 & 0.048695 & 0.6867 & 0.8895 & 0.6401 & 0.079567
\\

&  MLP & 0.7892 & 0.8211 & 0.7870 & 0.023389 & 0.7920 & \textbf{0.9296} & 0.7462 & 0.013501 
\\

& KNN & 0.9022 & 0.8431 & \textbf{0.9803} & \textbf{0.000579} & 0.8169 & 0.8944 & 0.7773 & \textbf{0.005129} 
\\
\bottomrule
\end{tabularx}
\end{adjustwidth}
\end{table}

\subsection{Analysis of Overfitting}
\label{sec:overfitting}

\begin{figure}[H]
\isPreprints{}
\begin{adjustwidth}{-\extralength}{0cm}
\centering
   \includegraphics[width=0.4\linewidth]{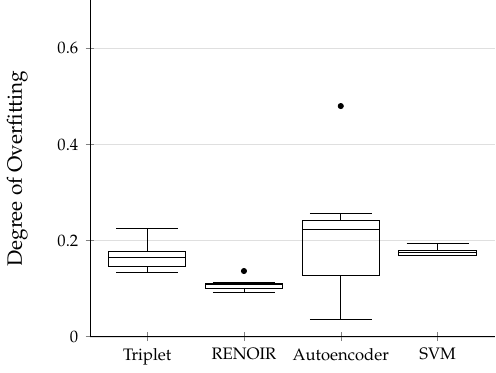}%
\hspace{0.05\linewidth}
   \includegraphics[width=0.4\linewidth]{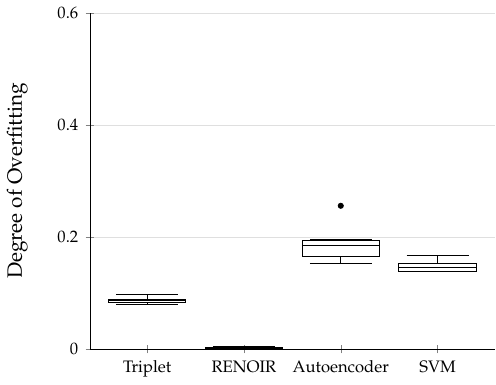}%
\end{adjustwidth}\vspace{-5pt}
\caption{Box plot showing degree of overfitting of the proposed triplet network and baseline models in binary classification. The relative difference in macro F1 score of a model between the train and test data splits was used to measure the generalisation gap as a proxy for overfitting. Models were trained on the Lycos2017 dataset with (\textbf{Left}): 10 attack samples per class in the training dataset and (\textbf{Right}): 160 attack samples per class in the training dataset. \label{fig:overfitting_binary}}
\end{figure}

\begin{figure}[H]
\isPreprints{}
\begin{adjustwidth}{-\extralength}{0cm}
\centering
   \includegraphics[width=0.4\linewidth]{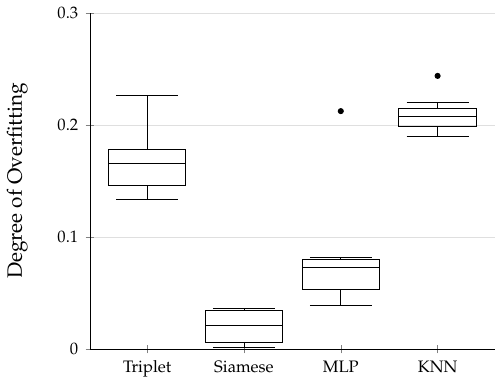}%
\hspace{0.05\linewidth}
   \includegraphics[width=0.4\linewidth]{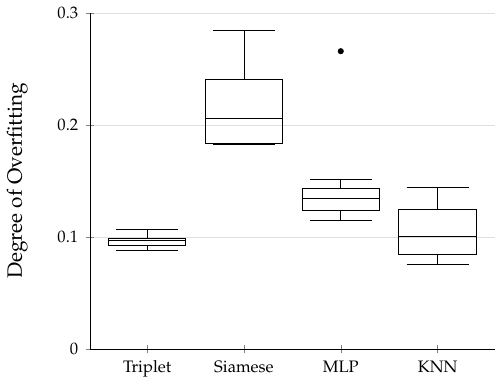}%
\end{adjustwidth}\vspace{-5pt}
\caption{Box plot showing degree of overfitting of the proposed triplet network and baseline models in multiclass classification. The relative difference in macro F1 score of a model between the train and test data splits was used to measure the generalisation gap as a proxy for overfitting. Models were trained on the Lycos2017 dataset with (\textbf{Left}):  10 attack samples per class in the training dataset and (\textbf{Right}): 160 attack samples per class in the training dataset.\label{fig:overfitting_multiclass}}
\end{figure}

The degree of overfitting experienced by the triplet network is compared to that of baseline models in Figure~\ref{fig:overfitting_binary} for binary classification and in Figure~\ref{fig:overfitting_multiclass} for multiclass classification. The relative difference in macro F1 score between the training and test data splits was used to measure the generalisation gap, which serves as a proxy for overfitting.

\vspace{-9pt}

Overfitting is generally characterised by a discrepancy between training and test performance, arising when a model learns patterns that do not generalise beyond the training data. In the few-shot regime considered in this work, this effect is fundamentally driven by the limited number of training samples. With very small datasets, the empirical training distribution provides only a noisy estimate of the underlying data-generating process. As a result, models may fit patterns that are specific to the sampled training data and do not generalise to unseen data. Therefore, in this setting, the observed overfitting is largely a consequence of undersampling the underlying data distribution.

In the binary classification setting, contrastive learning demonstrated greater robustness to this generalisation gap, with both RENOIR and the triplet network exhibiting the smallest discrepancy between training and test performance. Compared to traditional anomaly detection methods, the triplet network achieved a lower median gap than both the SVM and autoencoder, regardless of whether it was trained on 10 or 160 malicious samples per class. Notably, when trained with 160 samples per class, the triplet network showed the lowest generalisation gap in terms of both median and variance.

In the multiclass classification setting, the triplet network demonstrated reduced overfitting compared to KNN, regardless of whether it was trained with 10 or 160 malicious samples per class. This effect is especially pronounced at 160 samples per class, where the triplet network exhibits both a lower median overfitting value and significantly less variance than all other baseline models. This reduction in overfitting helps explain the observed performance gains when applying KNN to the triplet network’s learned embeddings, as opposed to using the original feature space. By mitigating overfitting, the triplet network's generalisation ability proves well-suited for scenarios with limited training data: a common constraint in practical NIDS applications.

\section{Ablations}
\label{sec:Ablations}

In this section the impact of various design choices is evaluated through ablation studies. Section~\ref{subsec:ablations_siamese_to_triplet} begins by offering a direct comparison between the triplet and Siamese networks when trained using an identical procedure. Section~\ref{subsec:ablations_mining_algorithms} continues the analysis of triplet networks by comparing the performance of various triplet mining algorithms. Section~\ref{subsec:ablations_distance_metric} compares the performance of various distance metrics when optimised by the triplet loss function and Section~\ref{subsec:ablations_infernce_algos} compares the performance of various inference algorithms when paired with the triplet network. Experiments in this section follow the experimental procedure described in Section~\ref{sec:evaluation_proceedure}, with the exception of Section~\ref{subsec:benign_samples_effect} and Section~\ref{subsec:inference_class_imbalance_effect}, which explore the effects of the the number of benign samples in the training set and class balancing techniques during inference, respectively. In these sections, the optimised models found during the hyperparameter search in Section~\ref{sec:experiments} were instead used to isolate the impacts of these factors.

\subsection{Comparison to Siamese Networks}
\label{subsec:ablations_siamese_to_triplet}

\begin{figure}[H]
\isPreprints{\centering}{}

\includegraphics[]{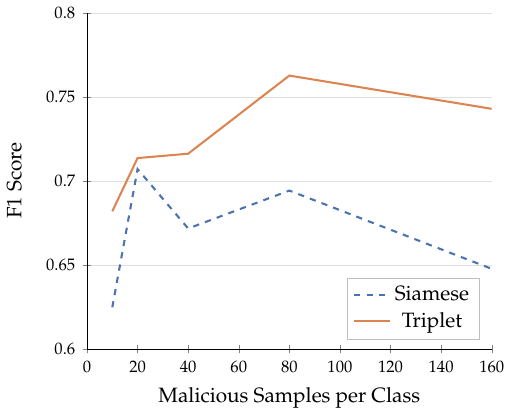}\vspace{-4pt}
\caption{Comparison of the Siamese network with the triplet network as the number of malicious samples per class in the training set is varied. The triplets used to train the triplet network are randomly sampled, following the methodology of previous works.\label{fig:siamese_triplet_comparison}}
\end{figure}

To enable a direct comparison between Siamese and triplet networks, the triplet network was trained using the procedures outlined in prior work~\cite{hanan_developing_10.115/3437984.3458842}. Specifically, it was trained on a fixed set of 30,000 triplets sampled prior to training, and inference was conducted by measuring the distance between a test sample's embedding and the embeddings of randomly selected training samples from each class. The macro F1 scores for both models, across varying training set sizes, are shown in Figure~\ref{fig:siamese_triplet_comparison}. The triplet network outperforms the Siamese network at all dataset sizes, with its performance improving more consistently as training size increases. This trend suggests that the triplet network maintains greater precision as recall improves, making it more effective in data-scarce scenarios.
\vspace{-6pt}

\unskip

\subsection{Comparison of Triplet Mining Algorithms}
\label{subsec:ablations_mining_algorithms}

The performance and convergence speed of the proposed triplet network are improved through the use of batch-all triplet mining; however, alternative triplet mining algorithms have been adopted in the previous literature. Batch-hard triplet mining seeks to accelerate training by focusing learning on the most difficult triplets. For each anchor in the batch, a triplet is formed by selecting the furthest positive embedding and closest negative embedding within the batch, resulting in the triplet with the highest loss. The triplet loss is then calculated as shown in Equation~(\ref{eq:batch_hard_triplet_loss})~\cite{in_defence_of_triplet_loss}.

\begin{linenomath}
\begin{equation}
L_{BH}(x, y) = \frac{1}{B} \sum_{a = 0}^{B-1} \; \underset{\substack{0 \leq p < B \\ y_p = y_a \\ 0 \leq n < B \\ y_n \neq y_a}}{\max} L_{TL}(x_a, x_p, x_n)\label{eq:batch_hard_triplet_loss}
\end{equation}
\end{linenomath}
     
Batch semi-hard triplet mining instead calculates the triplet loss for the hardest triplet of each anchor under the additional constraint that the positive embedding must be closer to the anchor than the negative embedding, as shown in Equation~(\ref{eq:batch_semi_hard_triplet_loss}).

\begin{linenomath}
\begin{equation}
L_{BSH}(x, y) = \frac{1}{B} \sum_{a = 0}^{B-1} \; \underset{\substack{L_{TL}(x_a, x_p, x_n) < m \\ 0 \leq p < B \\ y_p = y_a \\ 0 \leq n < B \\ y_n \neq y_a}}{\max} L_{TL}(x_a, x_p, x_n)\label{eq:batch_semi_hard_triplet_loss}
\end{equation}
\end{linenomath}

\begin{figure}[H]

\isPreprints{\centering}{}

\includegraphics[]{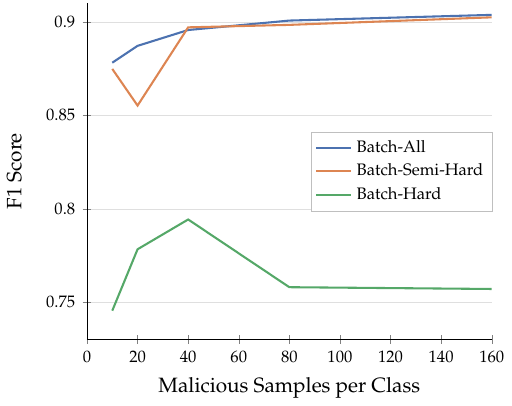}\vspace{-3pt}
\caption{Comparison of online-triplet-mining algorithms on the CICIDS2017 dataset when trained on varying amounts of malicious samples. Batch-all triplet mining (blue)
 is compared with batch semi-hard (orange) and batch hard triplet mining (green).\label{fig:comparison_of_triplet_mining_algorithms}}
\end{figure}

A comparison of each of the triplet mining algorithms is given in Figure~\ref{fig:comparison_of_triplet_mining_algorithms} across each training dataset size. Batch-hard mining consistently underperforms the other strategies on the CICIDS2017 dataset. Batch-all mining marginally outperforms batch semi-hard mining across most scenarios, suggesting that leveraging all available triplet information within a batch leads to more stable and effective learning.

\vspace{-3pt}
\unskip

\subsection{Comparison of Distance Metrics}
\label{subsec:ablations_distance_metric}

\begin{figure}[H]
\isPreprints{\centering}{} 

\includegraphics[]{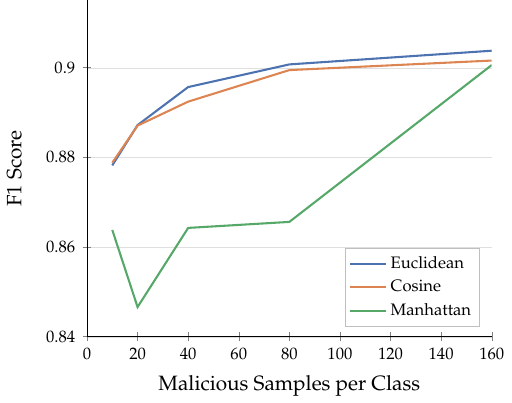}\vspace{-3pt}
\caption{Comparison of distance metrics when used to train the triplet network. The Euclidean distance (blue) is compared to the cosine distance (orange) and the Manhattan distance (green).}\label{fig:comparison_of_distance_metrics}
\end{figure}

In the proposed approach, the triplet loss is used to optimise the embedded space using the Euclidean distance. Similarly, the loss function can be optimised using alternative distance metrics, for example, the Manhattan and cosine~\cite{dino} distance metric. Figure~\ref{fig:comparison_of_distance_metrics} compares the macro F1 scores of triplet networks trained using Euclidean, cosine, and Manhattan distance metrics across varying quantities of malicious training samples. The Euclidean distance consistently yields the highest performance, slightly outperforming cosine distance across all training set sizes. In contrast, the Manhattan distance performs poorly, particularly on smaller datasets.
\vspace{-3pt}
\unskip

\subsection{Comparison of Inference Algorithm}
\label{subsec:ablations_infernce_algos}

While the KNN classifier was effective in this work, several alternative approaches have been explored in existing research. These include modifications of the KNN, such as a soft-voting KNN which assigns a probability to a test sample's class membership by summing the distances of a test embedding to the neighbours of each class. Additionally, a weighted version of KNN applies a temperature-scaled Boltzmann--Gibbs distribution to emphasise closer embeddings~\cite{dino}. Some NIDS approaches forego neighbour searches entirely, comparing test samples to randomly sampled training instances from each \mbox{class~\cite{hanan_leveraging_DBLP:journals/corr/abs-2006-15343,hanan_developing_10.115/3437984.3458842}}.

\begin{figure}[H]
\isPreprints{\centering}{} 

\includegraphics[]{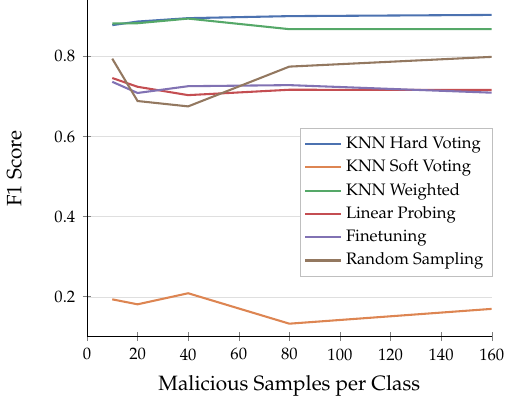}\vspace{-3pt}
\caption{Comparison of inference algorithms to predict labels using a trained model. \label{fig:comparison_of_inference_algorithms}}
\end{figure}

Another common approach to inference after training a contrastive learning model is to learn decision boundaries using gradient descent. Linear probing involves freezing the model's weights before training a linear classification head to perform classification. Similarly, fine-tuning jointly trains the pretrained model and classification head. In Figure~\ref{fig:comparison_of_inference_algorithms} each of the aforementioned inference algorithms is evaluated and compared across a variety of training dataset sizes. A hard-voting KNN was found to be the most effective algorithm.

\unskip

\subsection{Effect of the Number Benign Samples During Training}
\label{subsec:benign_samples_effect}

To assess the effect of benign class size during training, the number of benign samples included in the CICIDS2017 training set was varied while keeping the number of malicious samples fixed to 160 per class. The results for binary classification are reported in Table~\ref{tab:benign_sample_ablation}. Across all benign sample sizes, the proposed triplet network achieved the highest macro F1 score and precision and lowest false-positive rate. This indicates that the learned representation remains effective even when the quantity of benign training data is varied.

The results also suggest that increasing the number of benign samples beyond 5000~provides only marginal improvement for the triplet network. In particular, macro F1 remains stable across all settings, while the false-positive rate decreases as more benign samples are made available. This is consistent with the use of the KNN classifier at inference time, where a larger benign reference set provides a more accurate estimate of the benign class distribution and therefore improves robustness against false positives.

In contrast, the baseline models show greater sensitivity to the number of benign samples. The autoencoder performs poorly when only 1000 or 5000 benign samples are available, but improves substantially when 10,000 or more benign samples are used. RENOIR achieves relatively strong recall across all settings, but consistently exhibits lower precision and higher false-positive rates than the triplet network. Overall, these results suggest that the proposed triplet network is more robust to reductions in benign training data while maintaining strong binary classification performance.

\begin{table}[H]
\centering
\caption{Sensitivity of binary classification performance on CICIDS2017 to the number of benign samples used in the training data. Bold values indicate the best performance for each metric, with lower values preferred for false-positive rate}.
\label{tab:benign_sample_ablation}
\begin{tabularx}{\textwidth}{lCCCCC}
\toprule
\textbf{Benign Samples} & \textbf{Model} & \textbf{Macro F1} & \textbf{Recall} & \textbf{Precision} & \textbf{FP Rate} \\
\midrule
\multirow{3}{*}{1000}
& Triplet & \textbf{0.9504} & 0.9330 & \textbf{0.9704} & \textbf{0.004624} \\
& RENOIR  & 0.9132 & \textbf{0.9441} & 0.8892 & 0.027142 \\
& AE      & 0.5225 & 0.6880 & 0.5734 & 0.424422 \\
\midrule
\multirow{3}{*}{5000}
& Triplet & \textbf{0.9566} & \textbf{0.9283} & \textbf{0.9901} & \textbf{0.000447} \\
& RENOIR  & 0.8754 & 0.8855 & 0.8857 & 0.024076 \\
& AE      & 0.5323 & 0.6685 & 0.5662 & 0.405863 \\
\midrule
\multirow{3}{*}{10,000}
& Triplet & \textbf{0.9548} & 0.9244 & \textbf{0.9913} & \textbf{0.000161} \\
& RENOIR  & 0.9307 & \textbf{0.9451} & 0.9192 & 0.018069 \\
& AE      & 0.8440 & 0.8031 & 0.9117 & 0.011353 \\
\midrule
\multirow{3}{*}{20,000}
& Triplet & \textbf{0.9532} & \textbf{0.9216} & \textbf{0.9915} & \textbf{0.000064} \\
& RENOIR  & 0.8620 & 0.8617 & 0.8850 & 0.016363 \\
& AE      & 0.8465 & 0.8075 & 0.9116 & 0.011984 \\
\bottomrule
\end{tabularx}
\end{table}

\subsection{Effect of Class Imbalance During Inference}
\label{subsec:inference_class_imbalance_effect}

To validate the claim that the use of KNN at inference time reintroduces the class imbalance present in the training distribution, we compared the default KNN classifier operating on the unbalanced reference set with several alternative inference procedures. Specifically, we evaluated: (i) KNN using the original unbalanced reference set, (ii) KNN using a class-balanced reference set, (iii) a linear probe trained on the learned embeddings, and (iv) an imbalanced linear classifier trained on the embeddings. The results are shown in Table~\ref{tab:inference_class_imbalance_ablation} for multiclass classification on the CICIDS2017 dataset at 160 samples per malicious~class.

Across all training set sizes, the default KNN classifier on the unbalanced reference set achieved the highest macro F1, recall, and precision. In particular, it consistently maintained very high precision while also preserving a very low false-positive rate. This provides empirical support for the claim that retaining the original class imbalance during inference is beneficial in the binary intrusion detection setting, where benign traffic dominates and false positives must be minimised.

\begin{table}[H]
\centering
\caption{Sensitivity of classification performance on CICIDS2017 to the use of class-rebalancing during inference.Bold values indicate the best performance for each metric, with lower values preferred for false-positive rate}.
\label{tab:inference_class_imbalance_ablation}
\begin{tabularx}{\textwidth}{lcCCCC}
\toprule
\textbf{Attack Samples} & \textbf{Inference} & \textbf{Macro F1} & \textbf{Recall} & \textbf{Precision} & \textbf{FP Rate} \\
\midrule
\multirow{4}{*}{10}
& Imbalanced Linear & 0.1900 & 0.2000 & 0.1810 & \textbf{0.000005} \\
& Linear Probe      & 0.5048 & 0.6383 & 0.5064 & 0.190283 \\
& Balanced KNN      & 0.3779 & 0.4672 & 0.3989 & 0.271022 \\
& KNN               & \textbf{0.8602} & \textbf{0.7878} & \textbf{0.9638} & 0.000251 \\
\midrule
\multirow{4}{*}{20}
& Imbalanced Linear & 0.2603 & 0.2624 & 0.2632 & \textbf{0.000076} \\
& Linear Probe      & 0.4788 & 0.6437 & 0.4722 & 0.237753 \\
& Balanced KNN      & 0.4307 & 0.5627 & 0.4392 & 0.197239 \\
& KNN               & \textbf{0.8760} & \textbf{0.8019} & \textbf{0.9837} & 0.000333 \\
\midrule
\multirow{4}{*}{40}
& Imbalanced Linear & 0.3284 & 0.3246 & 0.3367 & 0.000301 \\
& Linear Probe      & 0.4944 & 0.6196 & 0.5338 & 0.281856 \\
& Balanced KNN      & 0.4375 & 0.5825 & 0.4637 & 0.231805 \\
& KNN               & \textbf{0.8927} & \textbf{0.8225} & \textbf{0.9902} & \textbf{0.000157} \\
\midrule
\multirow{4}{*}{80}
& Imbalanced Linear & 0.3008 & 0.2964 & 0.3575 & \textbf{0.000171} \\
& Linear Probe      & 0.5525 & 0.6928 & 0.5834 & 0.195583 \\
& Balanced KNN      & 0.5150 & 0.6175 & 0.5039 & 0.108013 \\
& KNN               & \textbf{0.9019} & \textbf{0.8389} & \textbf{0.9867} & 0.000265 \\
\midrule
\multirow{4}{*}{160}
& Imbalanced Linear & 0.3794 & 0.3755 & 0.4544 & 0.002183 \\
& Linear Probe      & 0.4619 & 0.6380 & 0.4983 & 0.298481 \\
& Balanced KNN      & 0.3359 & 0.4465 & 0.3926 & 0.379965 \\
& KNN               & \textbf{0.8041} & \textbf{0.7538} & \textbf{0.8853} & \textbf{0.000440} \\
\bottomrule
\end{tabularx}
\end{table}

Using a balanced class-balanced reference set for KNN inference led to substantially worse performance across all metrics. Although this rebalancing might be expected to improve sensitivity to the minority class, in practice it reduced both precision and recall while greatly increasing the false-positive rate. This suggests that removing the natural class skew from the reference set weakens the ability of the classifier to exploit the true marginal class distribution during prediction.

The linear classifiers also performed poorly relative to KNN. The imbalanced linear classifier achieved the lowest false-positive rates for the smallest training sets, but this came at the cost of extremely poor recall and macro F1, indicating that it was overly conservative and failed to detect a large proportion of malicious samples. The standard linear probe achieved higher recall than the imbalanced linear classifier, but produced substantially higher false-positive rates and lower precision than the default KNN. Overall, these results support the interpretation given in Section~\ref{subsec:method_balanced_sampling}: the proposed model benefits from balanced training batches for representation learning, while preserving class imbalance during inference allows the KNN classifier to recover the true class prior and maintain a more favourable precision--recall--false positive trade-off.

\section{Discussion and Conclusions}
\label{sec:discussion_and_conclusion}

This work demonstrates the effectiveness of a triplet network trained using online triplet mining for practical network intrusion detection in a few-shot regime, where only a limited number of malicious samples per class are available. Through the use of contrastive learning, the model is able to effectively learn a discriminative embedded representation of the data which is resistant to class imbalance and undersampled class distributions during training. Additionally, by utilising a KNN classifier for inference, the marginal distribution of the classes is exploited, allowing the proposed model to effectively identify malicious traffic whilst minimising false-positive rates. 

The triplet network was initially evaluated in few-shot binary classification, where it was compared to baseline models across the anomaly detection and contrastive learning literature, including autoencoders, one-class SVMs, and Siamese networks. The triplet network outperformed all baseline models in macro F1 score on both the CICIDS2017 and Lycos2017 datasets when trained on a dataset containing at least 20 malicious samples per class. Although the triplet network was marginally surpassed by RENOIR when trained on the Lycos2017 dataset with only 10 samples per class, the triplet network achieved significantly lower false-positive rates across all training datasets, making it a more practical~approach.

In few-shot multiclass classification, the triplet network was compared to various multiclass classifiers, including a Siamese network, representing previous contrastive learning approaches, and an MLP, representing a traditional classification approach. Additionally, a KNN classifier was used to compare the embedded representation of the data learned by the triplet network with the original feature space. The triplet network was found to outperform all baseline models on macro F1 score across all training dataset sizes on the Lycos2017 dataset, and when trained on at least 40 malicious samples per class on the CICIDS2017 dataset. This disparity can likely be attributed to the relative simplicity of the CICIDS2017 dataset, which contains fewer features and classes than the Lycos2017 dataset, suggesting that the proposed triplet network is suited to more complex, realistic scenarios.

By improving the few-shot classification performance of network intrusion detection systems, this work aims to reduce the deployment delay experienced by the current generation of intrusion detection systems which occurs while malicious data is collected and annotated after the establishment of a new network or emergence of a zero-day attack. Furthermore, by providing effective classification with a low false-positive rate, the proposed triplet network provides a viable alternative to anomaly-detection-based~approaches. 

Despite these promising results, several limitations remain. First, evaluation was conducted on two widely used benchmark datasets, and further validation on more recent or diverse datasets, such as IoT or encrypted traffic scenarios, would provide a more comprehensive assessment of generalisability. Second, while this work focuses on contrastive and anomaly detection baselines relevant to practical NIDSs, it does not include comparisons to broader meta-learning approaches such as prototypical networks or model-agnostic meta-learning. These methods were excluded because their typical episodic training procedures and support/query data assumptions differ from the fixed-reference inference setting considered in this work, making a controlled comparison beyond the scope of the present study. Such approaches therefore represent an important direction for future work.

The proposed approach is also naturally extensible to incremental and streaming intrusion detection settings. As inference is performed using a KNN classifier over learned embeddings, new attack classes can be incorporated by simply adding newly labelled samples to the reference set, without requiring full retraining of the model. This enables rapid adaptation to emerging threats. In streaming environments, this could be combined with periodic model updates and memory management strategies to maintain an efficient and representative reference set.

Finally, while the use of KNN at inference time provides strong performance and allows the model to exploit class imbalance, it introduces additional computational overhead that may impact latency in real-time deployment scenarios. This could be mitigated through the use of approximate nearest neighbour search methods, which provide sublinear query time while maintaining high retrieval accuracy, making them well suited for deployment in large-scale or high-throughput NIDS environments.

\appendixtitles{yes}
\appendixstart
\appendix

\section{Hyperparameter Settings}
\label{app:hyperparameters}

Table~\ref{tab:seed_configuration} lists the random seeds used to control the experimental procedure. Randomisation was controlled using the Python \texttt{random} module, which was used to select the hyperparameters for each model, select train--test data splits, generate data subsets for each iteration of the tuning procedure and generate random seeds, which were passed to scikit-learn cross-validation functions via \texttt{random\_state}. The base seeds were randomly generated once, stored in the configuration files, and reused throughout the study for reproducibility. For each of the five repeated evaluations used for each model, dataset, and task, the subset sampling and hyperparameter seeds were incremented by one to generate distinct, but reproducible, data subsets and hyperparameter search runs.

\begin{table}[H]
   \centering
   \caption{Base random seeds used to initialise the experimental pipeline.}
   \label{tab:seed_configuration}
   \begin{tabularx}{\textwidth}{lC}
   \toprule
   \textbf{Property} & \textbf{Seed} \\
   \midrule
   Configuration & 0 \\
   Subset sampling & 19,048 \\
   Cross-validation split & 19,324 \\
   Hyperparameter search & 4564 \\
   Dataset split & 39,058,032 \\
   \bottomrule
   \end{tabularx}
\end{table}

Hyperparameters for gradient-based models were selected using random search, while the hyperparameters of gradient-free models were selected using exhaustive search. The full hyperparameter search spaces are reported in Table~\ref{tab:hyperparameter_search_spaces}. The software environment used for the experiments comprised Python 3.10.12, scikit-learn 1.7.2, and PyTorch 2.9.0+cu128. The code used for this work is stored in the repository available at \url{https://github.com/jackwilkie/few_shot_nids_triplet_mining} (accessed on 2 May 2026).

\begin{table}[H]
   \centering
   \caption{Hyperparameter search spaces used during model selection.}
   \label{tab:hyperparameter_search_spaces}
   
   \begin{tabularx}{\textwidth}{lCC}
   \toprule
   \textbf{Model} & \textbf{Parameter} & \textbf{Search Range} \\
   \midrule
   \multirow{5}{*}{All gradient-based models} & learning rate & \(\text{LogUniform}(10^{-6}, 10^{-3})\)\\
   & batch size & \(\{32, 64, 128, 256, 512, 1024\}\) \\
   & weight decay & \(\text{LogUniform}(10^{-6}, 0.05)\) \\
   & \(\beta_1, \beta_2\) & \(0.9, 0.999\) \\
   & epochs & \(200\) \\
   \midrule
   \multirow{3}{*}{All MLP-based models} & neurons & \(\{32, 64, 128, 256, 512, 1024\}\)\\
   & depth & \(\{1, 2, 3, 4\}\)\\
   & dropout & \(\{0.1, 0.2, 0.3\}\) \\
   \midrule
   \multirow{2}{*}{All contrastive models} & \(f_{o}\) & \(8, 16, 32, 64, 128\) \\
   & \(m\) & \(\{0.1, 0.2, \dots, 1.0\}\)\\
   \midrule
   All autoencoders & bottleneck ratio & \(\{0.25, 0.5, 0.75\}\)\\
   \midrule
   DAE-LR & \(\lambda\) & \(\{0.1, 0.2, \dots, 1.0\}\)\\
   \midrule
   SVM & \(\nu\) & \(\{0.05, 0.10, \dots, 1.00\}\)\\
   \midrule
   KNN & \(k\) & \(\{1, 2, 4, 8, 16, 32, 64, 128\}\)\\
   \bottomrule
   \end{tabularx}
\end{table}

\authorcontributions{Conceptualisation, J.W., H.H., M.B., C.T. and R.A.; methodology, J.W., H.H., C.T., and R.A.; software, J.W.; validation, J.W., H.H. and R.A.; formal analysis, J.W.; investigation, J.W.; resources, R.A.; data curation, J.W., and H.H.; writing---original draft preparation, J.W.; writing---review and editing, J.W., H.H., C.T., M.B., and R.A.; visualisation, J.W.; supervision, R.A., C.T. and H.H.; project administration, C.T. and R.A.; funding acquisition, R.A. All authors have read and agreed to the published version of the manuscript.}

\funding{This research received no external funding.}

\institutionalreview{Not applicable.} 

\informedconsent{Not applicable.}

\dataavailability{The data presented in this study are available in GitHub at \url{https://github.com/jackwilkie/few_shot_nids_triplet_mining} (accessed on 2 May 2026).
 These data were derived from the following resources available in the public domain: CICIDS2017 {Dataset: }\url{https://www.unb.ca/cic/datasets/ids-2017.html} (accessed on 2 May 2026), Lycos2017 {Dataset:} \url{https://lycos-ids.univ-lemans.fr/} (accessed on 2 May 2026).}

\conflictsofinterest{The authors declare no conflicts of interest.} 

\begin{adjustwidth}{-\extralength}{0cm}

\reftitle{References}

\PublishersNote{}
\end{adjustwidth}
\end{document}